\documentclass[
 superscriptaddress,
 reprint,
 amsmath, amssymb,
 aps, preprintnumbers, nofootinbib, floatfix
]{revtex4-1}

\usepackage{graphicx}
\usepackage{dcolumn}
\usepackage{bm}
\usepackage{color}
\usepackage[colorlinks=true,allcolors=Blue,pdfusetitle]{hyperref}
\usepackage{rotating}
\usepackage{multirow}
\usepackage{graphicx}
\usepackage[dvipsnames]{xcolor}
\usepackage{svg}
\usepackage{esdiff}

\newcommand{\orcid}[1]{\href{https://orcid.org/#1}{\includegraphics[width=10pt]{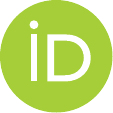}}}

\begin{document}
\preprint{N3AS-21-016}
\title{Towards Probing the Diffuse Supernova Neutrino Background in All Flavors}

\author{Anna M. Suliga \orcid{0000-0002-8354-012X}}
\email{asuliga@berkeley.edu}
\affiliation{
Niels Bohr International Academy and DARK, Niels Bohr Institute, \\
University of Copenhagen, Blegdamsvej 17, 2100, Copenhagen, Denmark
}
\affiliation{
Department of Physics, University of California Berkeley, Berkeley, California 94720, USA}
\affiliation{
Department of Physics, University of Wisconsin--Madison,
Madison, Wisconsin 53706, USA}

\author{John F. Beacom \orcid{0000-0002-0005-2631}}
\email{beacom.7@osu.edu}
\affiliation{
Center for Cosmology and AstroParticle Physics (CCAPP), Ohio State University, Columbus, OH 43210, USA
}
\affiliation{
Department of Physics,  Ohio State University,  Columbus,  OH 43210,  USA
}
\affiliation{
Department of Astronomy,  Ohio State University,  Columbus,  OH 43210,  USA
}

\author{Irene Tamborra \orcid{0000-0001-7449-104X}}
\email{tamborra@nbi.ku.dk}
\affiliation{
Niels Bohr International Academy and DARK, Niels Bohr Institute, \\
University of Copenhagen, Blegdamsvej 17, 2100, Copenhagen, Denmark
}

\date{25 January 2022}


\begin{abstract}
Fully understanding the average core-collapse supernova requires detecting the diffuse supernova neutrino background (DSNB) in all flavors.  While the DSNB $\bar{\nu}_e$ flux is near detection, and the DSNB $\nu_e$ flux has a good upper limit and prospects for improvement, the DSNB $\nu_x$ (each of $\nu_\mu, \nu_\tau, \bar{\nu}_\mu, \bar{\nu}_\tau$) flux has a poor limit and heretofore had no clear path for improved sensitivity.  We show that a succession of xenon-based dark matter detectors --- XENON1T (completed), XENONnT/LUX-ZEPLIN (running), and DARWIN (proposed) --- can dramatically improve sensitivity to DSNB $\nu_x$ the neutrino-nucleus coherent scattering channel.  XENON1T could match the present sensitivity of $\sim 10^3 \; \mathrm{cm}^{-2}~\mathrm{s}^{-1}$ per $\nu_x$ flavor, XENONnT/LUX-ZEPLIN would have linear improvement of sensitivity with exposure, and a long run of DARWIN could reach a flux sensitivity of $\sim 10 \; \mathrm{cm}^{-2}~\mathrm{s}^{-1}$.  Together, these would also contribute to greatly improve bounds on non-standard scenarios.  Ultimately, to reach the standard flux range of $\sim 1 \; \mathrm{cm}^{-2}~\mathrm{s}^{-1}$, even larger exposures will be needed, which we show may be possible with the series of proposed lead-based RES-NOVA detectors.
\end{abstract}

\maketitle


\section{Introduction}
\label{sec:IntroductionDSNB}

The most basic prediction about the neutrino emission from core collapse supernovae is that there is always a huge release of energy (approximately $2 \times 10^{59}$~MeV), shared comparably among all flavors.  {\it But this prediction is untested.}  The answer is critical to our understanding of core collapse, supernova nucleosynthesis, neutrino properties, and tests of new physics~\cite{Jegerlehner:1996kx, Duan:2010bg, Raffelt:2011nc, Burrows:2012ew, Adams:2013ana, Mirizzi:2015eza, Janka:2016fox, Martinez-Pinedo:2017ksl, Horiuchi:2018ofe, Warren:2019lgb, Li:2020ujl, Burrows:2020qrp, Tamborra:2020cul, Nunokawa:1997ct, Hidaka:2006sg, Fischer:2010wp, Tamborra:2011is, Warren:2014qza, Arguelles:2016uwb, Fischer:2017lag, Rembiasz:2018lok, Xiong:2019nvw, Suliga:2019bsq, Syvolap:2019dat, Warren:2019lgb, Shalgar:2019rqe, Mastrototaro:2019vug, Suliga:2020vpz, Tang:2020pkp, Suliga:2020jfa, Fischer:2020xjl}.  With SN 1987A, only $\bar{\nu}_e$ events were observed~\cite{PhysRevLett.58.1490, Bionta:1987qt, ALEXEYEV1988209}, due to the difficulties of detecting the other flavors.  While we now have a suite of detectors capable of testing this most basic prediction~\cite{SNEWS:2020tbu, Scholberg:2017czd}, for almost 35 years we have had no other nearby core collapse.  And, even when the next one occurs, and its neutrino emission is measured precisely, we will still not be certain of how other core collapses work.  New astrophysics or physics may make them more complex, more varied, and more different among flavors than we expect~\cite{Nunokawa:1997ct, Hidaka:2006sg, Fischer:2010wp, Tamborra:2011is, Warren:2014qza, Arguelles:2016uwb, Fischer:2017lag, Rembiasz:2018lok, Xiong:2019nvw, Suliga:2019bsq, Syvolap:2019dat, Shalgar:2019rqe, Mastrototaro:2019vug, Suliga:2020vpz, Tang:2020pkp, Suliga:2020jfa, Fischer:2020xjl}.

The only feasible method to probe the average neutrino emission per core collapse is through detecting the diffuse supernova neutrino background (DSNB), which arises from all past core collapses~\cite{Beacom:2010kk, Lunardini:2010ab, Mirizzi:2015eza, Vitagliano:2019yzm}.  The flux upper limit for $\bar{\nu}_e$ from Super-Kamiokande (SK) is $2.7 \; \mathrm{cm}^{-2}~\mathrm{s}^{-1}$ ~\cite{Bays:2011si, Zhang:2013tua, Super-Kamiokande:2021acd}, and it is poised to make a first detection~\cite{Beacom:2003nk, Super-Kamiokande:2021acd} (which could be improved by other upcoming and proposed detectors~\cite{JUNO:2015zny, Jinping:2016iiq, Abe:2018uyc, Sawatzki:2020mpb}).  {\it But our understanding of core collapse depends on probing the DSNB in all flavors, which is hard.}  For $\nu_e$, the Sudbury Neutrino Observatory (SNO) has set a limit $19 \; \mathrm{cm}^{-2}~\mathrm{s}^{-1}$~\cite{Beacom:2005it, Aharmim:2020agi}, and the Deep Underground Neutrino Experiment (DUNE) should improve on this~\cite{Zhu:2018rwc, Moller:2018kpn}.  The weakest link is $\nu_x$, i.e., each of $\nu_\mu, \nu_\tau, \bar{\nu}_\mu, \bar{\nu}_\tau$.  For these, the limits are $\sim 10^3 \; \mathrm{cm}^{-2}~\mathrm{s}^{-1}$ per flavor~\cite{Lunardini:2008xd} (using neutrino-electron scattering in SK), a big improvement from the prior value of $\sim 10^7$ set using smaller detectors~\cite{1992APh.....1....1A}.  New ideas are needed to improve DSNB sensitivity to all flavors~\cite{Tabrizi:2020vmo}.

In this paper, we show that the sensitivity to DSNB $\nu_x$ can be greatly improved through careful analyses in future direct-detection dark-matter detectors.  Two developments make this possible.  The COHERENT experiment made the first detection of the coherent elastic scattering of neutrinos on nuclei (CE$\nu$NS) process, which is sensitive to all flavors of neutrinos and antineutrinos~\cite{Akimov:2017ade}.  And direct-detection experiments for dark matter are rapidly becoming much more sensitive~\cite{EDELWEISS:2011epn, 2012EPJC...72.1971A, TEXONO:2013hrh, DEAP-3600:2017uua, DarkSide:2018bpj, SuperCDMS:2018mne, LUX:2018akb, Aprile:2018dbl, PandaX-4T:2021bab}.  As these detectors search for nuclear recoils from dark-matter scattering, they will ultimately also become sensitive to nuclear recoils from neutrino scattering, reaching the ``neutrino floor" for dark-matter searches~\cite{ Vergados:2008jp, Strigari:2009bq, Billard:2013qya, Baudis:2013qla, Ruppin:2014bra, OHare:2016pjy, Boehm:2018sux}.

Our main calculations focus on the upgrade path from XENON1T~\cite{XENON:2017lvq} (fiducial mass 1~ton, completed) to XENONnT/LUX-ZEPLIN (fiducial mass 4~ton, running)~\cite{XENON:2020kmp, LZ:2019sgr} to DARWIN (DARk matter WImp search with liquid xenoN; fiducial mass 40~ton, proposed)~\cite{Aalbers:2016jon}.  We show that existing XENON1T data could set a limit of $\sim 10^3 \; \mathrm{cm}^{-2}~\mathrm{s}^{-1}$ per flavor (comparable to the SK limit) and that DARWIN will ultimately be able to probe down to $\sim 10 \; \mathrm{cm}^{-2}~\mathrm{s}^{-1}$.  Even limits would be important for probing core-collapse models with new astrophysics or physics.  To ensure detection, larger detectors are needed.  In an appendix, we show that this may be possible with proposed lead-based detectors RES-NOVA-1,-2,-3~\cite{Pattavina:2020cqc}.

The remainder of this paper is organized as follows.  In Sec.~\ref{sec:promise}, we show that the DSNB flux dominates other neutrino fluxes in a narrow energy range.  In Sec.~\ref{sec:challenges}, we show that detection is much more challenging than this flux dominance suggests.  But in Sec.~\ref{sec:resolution}, we show that even current data allow a useful limit and that there are promising ways forward with future detectors.  In Sec.~\ref{sec:conclusions}, we conclude.  In the appendices, we give additional details on our calculations of the DSNB flux, present a new limit on supernova $\nu_x$ emission from SN 1987A, and summarize our results for lead-based detectors.


\section{Promise: Multi-Flavor Detection of the DSNB}
\label{sec:promise}

In this section, we review the calculation of the DSNB signal flux (Sec.~\ref{sec:promise-DSNB}) and how it compares to other neutrino fluxes, which act as an irreducible background to DSNB detection (Sec.~\ref{sec:promise-backgrounds}).  The key message is that there is a narrow energy window ($\simeq$ 20--30 MeV) where the DSNB flux dominates.  Similar results have been found in prior work~\cite{Strigari:2009bq, Billard:2013qya, Baudis:2013qla, Ruppin:2014bra, OHare:2016pjy}.


\subsection{Calculation of the DSNB Signal}
\label{sec:promise-DSNB}

The fundamental goal of DSNB searches is to measure the average neutrino emission per core collapse~\cite{Beacom:2010kk, Lunardini:2010ab, Mirizzi:2015eza, Vitagliano:2019yzm}. This can be determined only by neutrino observations, whereas all other DSNB inputs can be measured through electromagnetic observations. At lowest order, the neutrino emission is expected to be independent of the progenitor mass. The most important correction is if there is a successful supernova (leading to a $\simeq 1.5M_\odot$ neutron star; NS) or a failed supernova (leading to a stellar-mass black hole; BH).  For BH formation, the neutrino flux is expected to be somewhat {\it larger}, and the spectrum somewhat {\it harder}, due to the increased mass of the collapsed core as well as increased accretion onto it.  The complementary fractions of these outcomes, $f_\mathrm{NS}$ and $f_\mathrm{BH}$, are not well known~\cite{Kochanek:2008mp, Lien:2010yb, Gerke:2014ooa, Sukhbold:2015wba, Ertl:2015rga, Adams:2016ffj, Adams:2016hit, Davies:2020iom}.

Here we carry out the DSNB modeling simply; for overviews tackling the astrophysical uncertainties affecting the signal, see Refs.~\cite{Horiuchi:2008jz, Lunardini:2012ne, Nakazato:2015rya, Kresse:2020nto, Horiuchi:2020jnc}.  Following Ref.~\cite{Moller:2018kpn}, we consider various sub-ranges of zero-age main sequence progenitor masses within $M =$ 8--125$M_\odot$ that lead to NS or BH outcomes.  We model the neutrino emission per core collapse using the time-integrated outputs from simulations from the Garching group (1D hydrodynamical simulations  with Boltzmann neutrino transport)~\cite{Garc:SN}. For NS outcomes from lower-mass supernovae, we use their 9.6-$M_\odot$ model; for NS outcomes from higher-mass progenitors, we use their 27-$M_\odot$ model; and for all BH outcomes, we use their 40-$M_\odot$ models with different accretion rate (a slow-forming BH model and a fast-forming BH model). For simplicity, we neglect the intermediate scenario of fallback supernovae~\cite{Zhang:2007nw} because of their lower rate~\cite{Ertl:2015rga}; we assume that this scenario falls in one of the two categories above.

The all-sky DSNB flux for a single neutrino flavor, with emission spectrum $F(E_\nu^{\prime},M)$, can then be calculated as
\begin{equation}
\begin{aligned}
\Phi (E_\nu) =& \frac{c}{H_{0}} \int_{8 M_{\odot}}^{125 M_{\odot}} dM \int_0^{z_{\max}} dz \frac{R_{\mathrm{SN}}(z, M)}{\sqrt{\Omega_M (1+z)^{3} + \Omega_\Lambda}} \\
& \times \left[ f_\mathrm{NS} F_\mathrm{NS} \left( E_{\nu}^{\prime}, M\right) + f_{\mathrm{BH}} F_\mathrm{BH} \left( E_{\nu}^{\prime}, M \right) \right] \,,
\end{aligned}
\label{eq:DSNB1}
\end{equation}
where the neutrino energy at emission, $E_\nu^{\prime}$, is related to that at detection, $E_\nu$, by $E_\nu^{\prime} = E_\nu (1 + z)$, with $z$ the redshift.  The supernova rate density (successful plus failed) for a given mass is $R_\mathrm{SN}(z,M)$.  We use standard values for the other inputs (speed of light $c$, Hubble constant $H_0$, dark matter fraction $\Omega_M$, and dark energy fraction $\Omega_\Lambda$)~\cite{Zyla:2020zbs}.  The resulting flux spectrum $\Phi (E_\nu)$ has units of [cm$^{-2}$\:s$^{-1}$\:MeV$^{-1}$].

The star-formation rate density from Ref.~\cite{Yuksel:2008cu} is
\begin{equation}
\label{eq:sfr}
\dot{\rho}_\star(z) \propto \left[(1+z)^{-3.4} + \left(\frac{1+z}{5000}\right)^{0.03}+\left(\frac{1+z}{9}\right)^{0.35}\right]^{1/10}\ .
\end{equation}
This leads to the supernova rate density via
\begin{equation}
\label{eq:SN_Rate_DSNB}
R_\mathrm{SN} (z, M) = \frac{(dN/dM) \dot \rho_{\star}(z)}{\int_{0.1M_\odot}^{125M_\odot} dM \; M \; (dN/dM)} \, , 
\end{equation}
where the initial mass function follows a Salpeter form, $dN/dM \propto M^{-2.35}$~\cite{1955ApJ...121..161S}, but punctuated as in Ref.~\cite{Moller:2018kpn} by step functions based on the NS versus BH outcomes.  When integrated over mass, we refer to the result as $R_\mathrm{SN}(z)$.

For our fiducial DSNB model, following Ref.~\cite{Moller:2018kpn}, we adopt $R_\mathrm{SN}(z=0) = 1.25 \times 10^{-4} \ \mathrm{Mpc}^{-3} \ \mathrm{yr}^{-1}$ and $f_\mathrm{BH} \simeq 20\%$ (slow-forming BH model). For our minimal DSNB model, we change these to 0.75 (the prefactor) and $f_\mathrm{BH} \simeq 10 \%$ (fast-forming BH model). For our maximal DSNB model, we change these to 1.75 and $f_\mathrm{BH} \simeq  40 \%$ (slow-forming BH model). These ranges are based on a relatively conservative assessment of input values from observations of $R_\mathrm{SN}(z=0)$~\cite{Smartt:2008zd, Li:2010kc, Horiuchi:2011zz, Botticella:2011nd, Mattila:2012, Taylor:2014rlo, Strolger:2015kra} and $f_\mathrm{BH}$~\cite{Kochanek:2008mp, Lien:2010yb, Gerke:2014ooa, Sukhbold:2015wba, Ertl:2015rga, Adams:2016ffj, Adams:2016hit, Davies:2020iom}.  A higher $f_\mathrm{BH}$ increases the high-energy tail of the DSNB~\cite{Lunardini:2009ya}.  The weighted effective total energy and average energy values for our DSNB models, which are comparable to those used in other calculations, are specified in App.~\ref{app:DSNB}.

\begin{figure}[t]
\centering
\includegraphics[width=0.99\columnwidth]{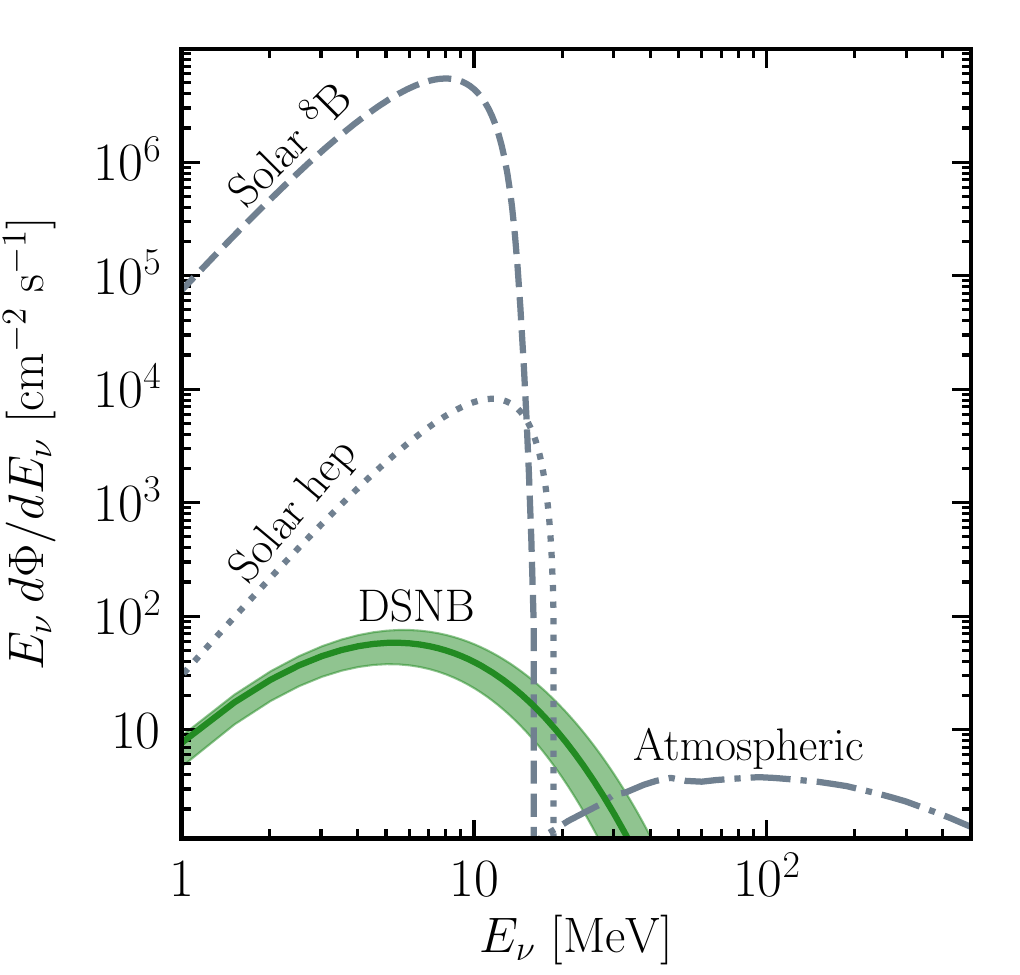}
\caption{
DSNB signal flux (sum of $\nu_\mu, \nu_\tau, \bar{\nu}_\mu, \bar{\nu}_\tau$) compared to the irreducible neutrino backgrounds (sums of all flavors).  The DSNB flux dominates only in a narrow energy window.
}
\label{fig:flux}
\end{figure}

Figure~\ref{fig:flux} shows the total DSNB $\nu_x$ spectrum (sum of $\nu_\mu, \nu_\tau, \bar{\nu}_\mu, \bar{\nu}_\tau$), with the line and band indicating our fiducial model and the range around it set by our minimal and maximal models.  
We show only the four $\nu_x$ flavors because the limits for $\bar{\nu}_e$ and $\nu_e$ are so much stronger.  We plot $E_\nu d\Phi/dE_\nu = (2.3)^{-1} d\Phi/d\log_{10}(E_\nu)$ to show the flux per logarithmic energy bin, matching the $x$-axis.  This model, when adapted to the $\bar{\nu}_e$ component, is consistent with constraints from SK~\cite{Super-Kamiokande:2021acd}. In App.~\ref{app:DSNB}, we provide more detail on the DSNB modeling and its uncertainties.  The other neutrino fluxes in the figure are explained in the next subsection.

The standard core-collapse supernova scenario predicts comparable fluxes for all neutrino flavors. But due to the large discrepancy between the limits on the $\nu_x$ and on other flavor components of the DSNB, we treat the $\nu_x$ emission as separate.  We neglect ordinary three-flavor mixing effects (including neutrino self-interactions~\cite{Lunardini:2012ne}) because if the $\nu_x$ flux  were really huge, this must be caused by unspecified new astrophysics or physics.  The latter might include active neutrino decays~\cite{Ando:2003ie, Fogli:2004gy, DeGouvea:2020ang}, non-standard interactions of DSNB with cosmic neutrino background~\cite{Goldberg:2005yw, Baker:2006gm, Farzan:2014gza, Jeong:2018yts} or sterile-neutrino mixing to the mu and tau sectors.  For the last, there might be a scenario where sterile neutrinos escape from the core and then decay to $\nu_x$~\cite{Dolgov:2000jw, Mastrototaro:2019vug}.

\subsection{Irreducible Neutrino Backgrounds}
\label{sec:promise-backgrounds}

Because the detection of CE$\nu$NS events in dark-matter detectors is flavor-blind and has only the nuclear recoil energy as an observable (e.g., no directionality), other neutrino fluxes form irreducible backgrounds to the DSNB signal. Other types of detector backgrounds, which will increase the difficulty of detecting the DSNB, are discussed in subsequent sections.

Figure~\ref{fig:flux} also shows these irreducible neutrino backgrounds. At low enough energies, solar-neutrino fluxes~\cite{Vinyoles:2016djt, Vitagliano:2019yzm} --- specifically from the $^8{\rm B}$ and $\mathrm{hep} ~(^3{\rm He + p})$ reactions, with endpoints of 15~MeV and $\simeq 19~\mathrm{MeV}$, respectively --- will overwhelm the main part of the DSNB spectrum. At high enough energies, atmospheric-neutrino fluxes (calculated for the location of the Gran Sasso National Laboratories, as appropriate for all the detectors considered here)~\cite{Battistoni:2002ew, Newstead:2020fie} will overwhelm the high-energy tail of the DSNB spectrum. For the solar- and atmospheric-neutrino fluxes, we use the sum of all flavors of neutrinos and antineutrinos, as appropriate for neutral-current detection. All other known neutrino fluxes are subdominant to those shown, and hence are neglected.  This figure is in good agreement with those in prior work~\cite{Strigari:2009bq, Billard:2013qya, Baudis:2013qla, Ruppin:2014bra, OHare:2016pjy}.


\section{Challenge: DSNB Detection in Dark-Matter Detectors}
\label{sec:challenges}

In this section, we review the challenges of DSNB detection in dark-matter detectors: the unfavorable differential cross section (Sec.~\ref{sec:challenges-CENNS}) and the detector size (Sec.~\ref{sec:challenges-detectors}).  The key message is that the DSNB signal that looked promising in Fig.~\ref{fig:flux} is now buried under backgrounds in Fig.~\ref{fig:rates} (see also Refs.~\cite{Strigari:2009bq, Billard:2013qya, Baudis:2013qla, Ruppin:2014bra, OHare:2016pjy}).  In the next section, we show a way forward.


\subsection{Coherent Elastic Neutrino-Nucleus Scattering}
\label{sec:challenges-CENNS}

In the CE$\nu$NS process~\cite{Freedman:1973yd}, a neutrino coherently and elastically interacts with a nucleus, causing it to recoil,
\begin{equation}
\label{eq:nucleus_scattering-DSNB}
\nu + A(Z, N) \rightarrow \nu + A(Z, N) \,, 
\end{equation}
where $A$ is the mass number, $Z$ the atomic number, and $N$ is the neutron number.  A coherent interaction requires the momentum transfer to be sufficiently low, leading to low recoil energies (typically in the keV range), making it very challenging to observe this reaction. For a neutrino of energy $E_\nu$ interacting with a nucleus of mass $m_A$, a range of nuclear recoil energies $E_r$ is produced, from zero up to a maximum, $E_r^\mathrm{max} = 2E_\nu^2 / (m_A + 2E_\nu) \simeq 2E_\nu^2 / m_A$, which corresponds to the configuration where the neutrino bounces straight backwards.  To induce xenon recoils of 1, 3, and 5~keV, the incoming neutrino energy must be larger than 8, 13.5, and 17.5~MeV, respectively.

Because this process is a neutral-current vector weak interaction, all active flavors of neutrino (and antineutrino) participate equally.  (We neglect an axial weak interaction contribution because in our work we focus on heavy nuclei, for which any axial contribution is suppressed by a factor of $\sim A$ ~\cite{Papoulias:2017qdn,Scholberg:2020pjn}.) The total cross section is large for a neutrino interaction, $\sigma_{\nu A} \simeq N^2 G_\mathrm{F}^2 E^2_\nu / 4 \pi$; for xenon, this is $\simeq 2.5 \times 10^{-41} (E_\nu/\mathrm{MeV})^2$ cm$^2$.  The differential cross section is~\cite{Freedman:1973yd}
\begin{equation}
\label{eq:sigma_nu_N-DSNB}
\frac{d\sigma_{\nu A}}{dE_r} \simeq \frac{G_\mathrm{F}^2 m_A}{4\pi} Q_w^2 \left(1 - \frac{m_A E_r}{2 E_\nu^2} \right) F^2(Q) \; \Theta (E_r^\mathrm{max} - E_r) \ ,
\end{equation}
where $Q_w = \left[N - Z(1 - 4\sin^2\theta_W)\right]$ is the weak vector nuclear charge, and the Helm-type form factor~\cite{Helm:1956zz}, which accounts for the partial loss of coherence when the momentum transfer ($Q = \sqrt{2 m_A E_r}$) is too large, is
\begin{equation}
\label{eq:form_factor-DSNB}
F(Q) = 3 \frac{j_1(Q R_0)}{Q R_0} \exp{\left(-\frac{1}{2} Q^2 s^2\right)} \ ,
\end{equation}
where $j_1$ is the first-order spherical Bessel function, the nuclear size is $R_0 = \sqrt{R^2 - 5s^2}$, $R = 1.2 A^{\frac{1}{3}}~\mathrm{fm}$, and $s \approx 0.5$ is the nuclear skin thickness~\cite{ENGEL1991114, Kozynets:2018dfo}.  The uncertainty associated with the form factor is less than 5\%~\cite{Scholberg:2005qs, Akimov:2017ade}.

\begin{figure}[t]
\centering
\includegraphics[width=0.99\columnwidth]{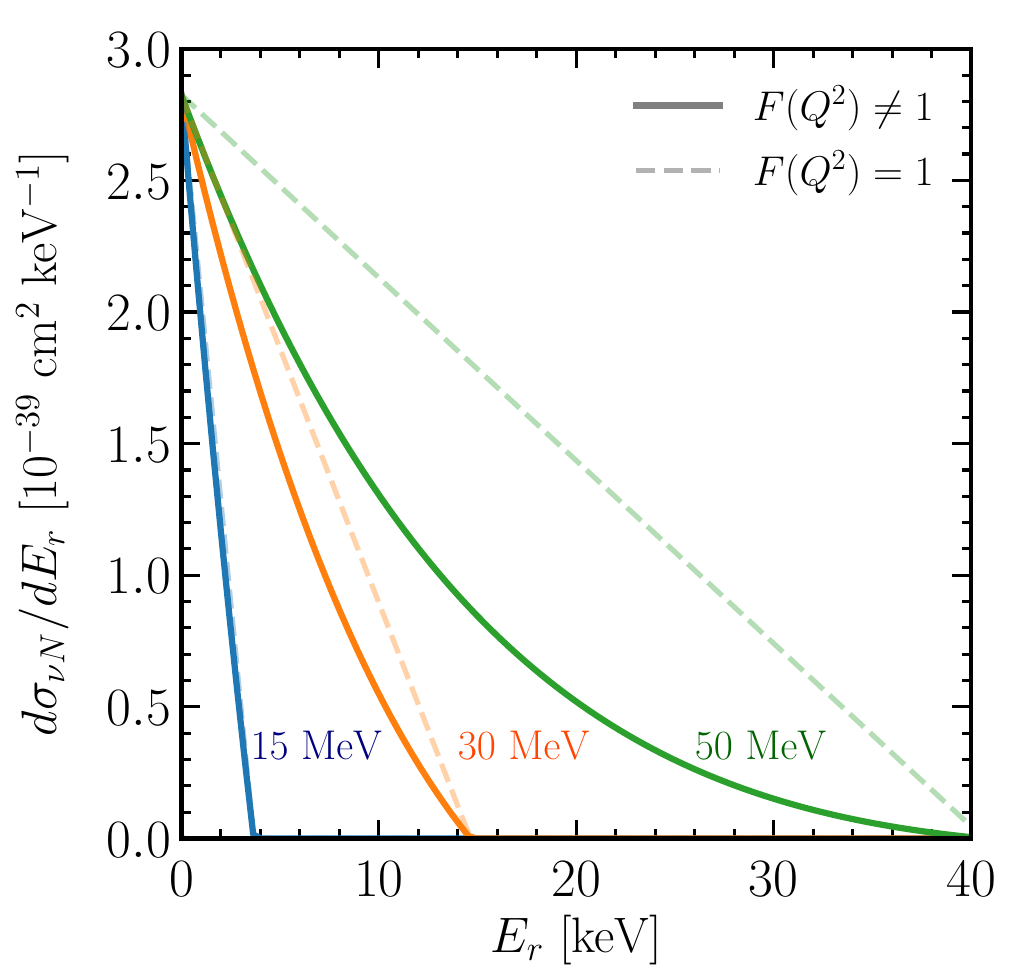}
\caption{
Differential cross section for the CE$\nu$NS process in terms of nuclear recoil energy, for xenon nuclei and three example neutrino energies.  The solid lines include the required nuclear form factor, whereas the dashed lines do not and are just for comparison.  The differential cross section favors low recoil energies, especially for increasing neutrino energy.
}
\label{fig:diff-cross-section}
\end{figure}

Figure~\ref{fig:diff-cross-section} shows the differential cross section, illustrating the points above.  A key challenge is that the smallest recoil energies are favored, but these are the hardest to detect.  With increasing neutrino energy, the maximum recoil energy increases, but then the form-factor suppression sharpens the preference for low recoil energies.


\subsection{Detection in Xenon-Based Detectors}
\label{sec:challenges-detectors}

Although the CE$\nu$NS cross section is large, detection is challenging because the recoil energies are low, the detectors that can measure them are small, and other neutrino fluxes are important.  In this section, we consider an idealized xenon detector, taking 1 ton as a fiducial mass (like XENON1T, but we defer most discussions of realistic details to the next section).  For comparison, the fiducial mass of SK, the leader in the search for DSNB $\bar{\nu}_e$, is 22,500 times larger.  Even taking into account the advantage of coherent scattering, xenon-based detectors will need to be much larger --- as planned --- to compete.

The working principle of a dual-phase xenon dark-matter detector is to separately detect prompt and delayed scintillation signals in the detector's liquid and gas regions~\cite{Aprile:2006kx}. The ratio between these signals allows for powerful discrimination between nuclear and electronic recoils, as required for dark-matter searches~\cite{XENON:2020kmp, Aalbers:2016jon}. For nuclear recoils compared to electronic recoils, the ionization density is higher, leading to faster recombination and a larger prompt scintillation signal (S1).  Most ionization electrons do not recombine, and instead drift in the applied electric field to the gas region, where they cause the delayed scintillation signal (S2). In the S1-S2 plane, signals from nuclear recoils then fall below those from electronic recoils~(see, e.g., Fig.~5 in Ref.~\cite{XENON:2020kmp}).

For simplicity, we work directly with the nuclear recoil energy, which can be estimated from a combination of S1 and S2 signals~\cite{Shutt:2006ed, Sorensen:2011bd}.  In the idealized case, the neutrino-induced recoil spectrum is
\begin{equation}
\label{eq:Event_rate_nu_N-DSNB}
\frac{dN_{\nu A}}{dE_r} = T N_t \int dE_\nu \ \frac{d\sigma_{\nu A}}{dE_r} \ \Phi(E_\nu)  \ ,
\end{equation}
where $T$ is the time of exposure, $N_t$ is the number of xenon nuclei, and $\Phi(E_\nu)$ is the neutrino flux spectrum. In our calculations, we include the weighting over xenon isotopes, though we suppress the notation.  Realistic detection effects are taken into account in the next section.

\begin{figure}[t]
\centering
\includegraphics[width=0.99\columnwidth]{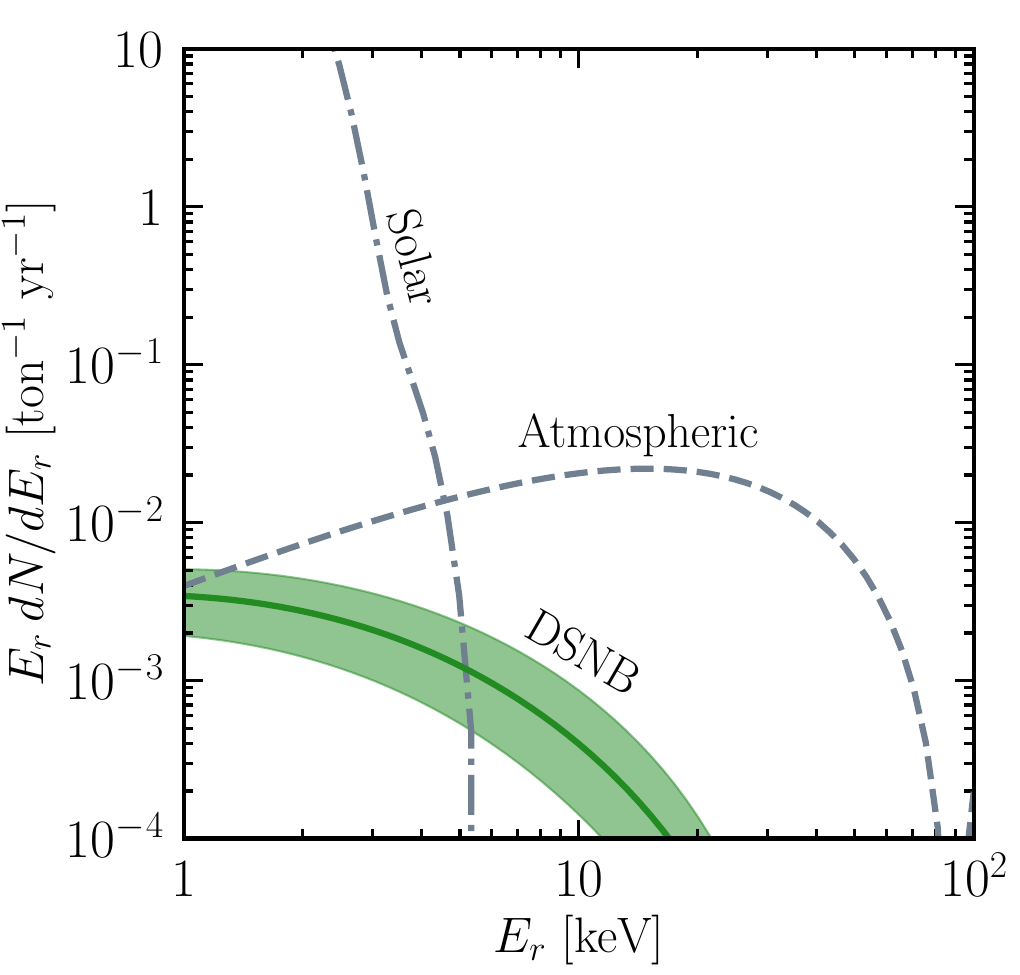}
\caption{
Expected nuclear-recoil spectrum for the DSNB (sum of $\nu_\mu, \nu_\tau, \bar{\nu}_\mu, \bar{\nu}_\tau$) and for various irreducible backgrounds, as labeled, corresponding to Fig.~\ref{fig:flux}.  Even in this idealized case, the DSNB would be very hard to detect.
}
\label{fig:rates}
\end{figure}

Figure~\ref{fig:rates} shows the DSNB signal in a 1 ton-year exposure of an ideal xenon detector.  With no information on the directionality of the events, other neutrino fluxes produce irreducible backgrounds (Sec.~\ref{sec:promise-backgrounds}), as shown.  Contrary to Fig.~\ref{fig:flux}, there is no longer an energy range where the DSNB signal dominates.  The reason is that signal events are pushed to lower energies by the differential cross section, and are thus obscured by the solar-neutrino background.  Further, the atmospheric-neutrino background pushes into the DSNB signal region.  In addition, present detectors are far too small to detect the DSNB, even if there were no backgrounds (for example, the rate within an energy range spanning a factor $e = 2.72$ can be estimated from the plot by multiplying the height of the curve by $\Delta \ln E = 1$).

\section{Resolution: Towards Future DSNB Sensitivity}
\label{sec:resolution}

In this section, we present our new results on how dark-matter detectors can place competitive limits on DSNB $\nu_x$.  In Sec.~\ref{sec:resolution-range}, we calculate the approximate range of the DSNB emission spectrum that can be probed.  In Sec.~\ref{sec:resolution-spectrum}, we calculate the nuclear recoil spectrum, here including realistic detector effects.  In Sec.~\ref{sec:resolution-statistics}, we calculate the DSNB flux sensitivity.


\subsection{Relevant Range of Neutrino Energy}
\label{sec:resolution-range}

\begin{figure}[b]
\centering
\includegraphics[width=0.99\columnwidth]{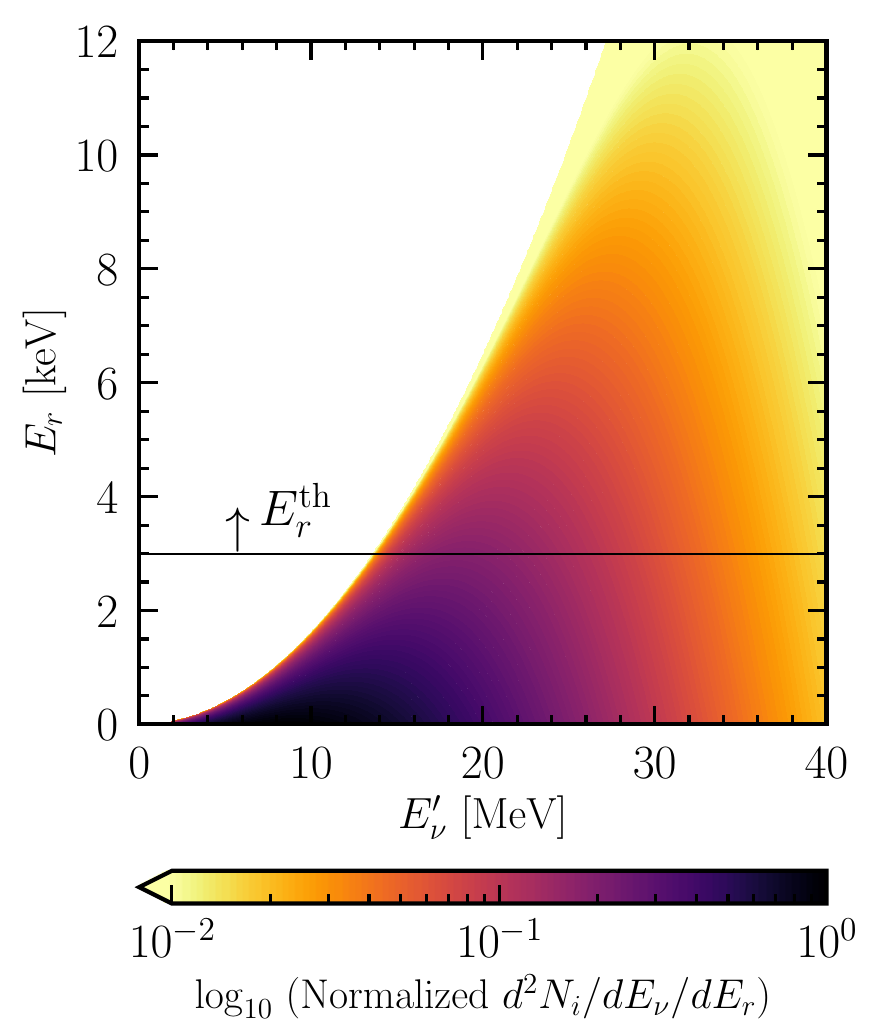}
\caption{
Relative weighting of detected DSNB $\nu_x$ events in an ideal xenon detector, in the plane of emitted neutrino energy and detected nuclear recoil energy.  The horizontal line shows a possible recoil-energy threshold.
}
\label{fig:Normalized_N}
\end{figure}

Figure~\ref{fig:Normalized_N} shows how the range of the nuclear recoil spectrum measured in the detector relates to the range of the DSNB spectrum probed at the source.  These results are calculated for an ideal xenon detector, as in the previous section.  For a fixed neutrino energy in emission, our calculation takes into account redshifting from the sources, as well as the weighting with the differential cross section.  For a fixed recoil energy, one can then read off the relevant range of neutrino energies at emission.  In anticipation of the next subsection, we indicate a possible recoil-energy threshold of 3 keV~\cite{Aalbers:2016jon}.  The relevant range for the neutrino emission energy is roughly 15--35 MeV, which is the tail of the spectrum.  The lower value is set by the detector response and the upper value is set by the falling DSNB spectrum.  This range is comparable to that for prior and future searches for DSNB $\bar{\nu}_e$~\cite{Super-Kamiokande:2021acd, Beacom:2003nk, Abe:2018uyc}, DSNB $\nu_e$~\cite{Beacom:2005it, Aharmim:2020agi, Moller:2018kpn, Zhu:2018rwc}, and DSNB $\nu_x$~\cite{Lunardini:2008xd}.  Typically, experiments note the range of detected (not emitted) energies they probe.  Correcting for this only strengthens the point that all DSNB searches only probe the tail of the emission spectrum.


\subsection{Predicted Nuclear Recoil Spectrum}
\label{sec:resolution-spectrum}

Starting in this subsection, we take into account several important detector effects that change the nuclear recoil spectrum in a xenon detector relative to the ideal case shown in Fig.~\ref{fig:rates}.  We model these effects based on the technical specifications achieved for the XENON1T detector.  For its successor detectors, we assume that these specifications will only improve, as they must meet the increasingly ambitious goals for dark-matter searches.

\begin{figure}[t]
\centering
\includegraphics[width=0.99\columnwidth]{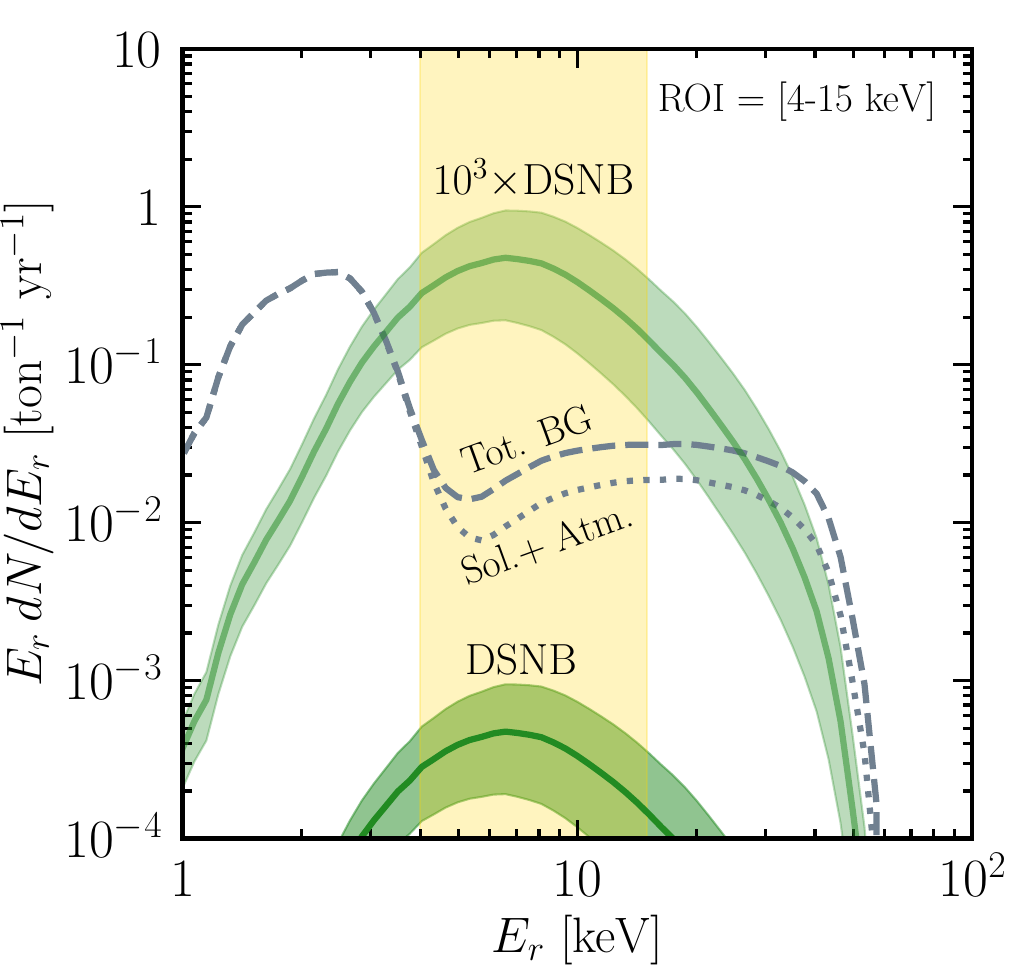}
\caption{
Same as Fig.~\ref{fig:rates}, but now taking into account realistic detector effects for XENON1T, including the neutron contribution to the total background.  Within the ROI, the DSNB spectra near the present limit dominate the backgrounds and are within reach of detection.  This sets the stage for improved sensitivity with larger exposures.
}
\label{fig:spectrum}
\end{figure}

Taking into account detector energy resolution has only modest effects, due to the narrow width of the smearing function ($\simeq$ 10--20\%, depending on energy)~\cite{Aprile:2018dbl, Schumann:2015cpa}.  To calculate the smeared spectrum, we finely bin the ideal recoil spectrum and convolve it with an energy-dependent Gaussian.  This moves events from the steep solar-neutrino background to higher energies by about 1 keV, but the other spectra are essentially unchanged.  The differential cross section causes much larger effects, though it only pushes events to lower energies, whereas detector energy resolution smears in both directions.

We then apply an energy-dependent detection efficiency~\cite{Aprile:2018dbl} to the smeared spectrum, which has large effects.  At energies below $\simeq 4$ keV, the low efficiency strongly suppresses event rates because of the requirement of having large enough S1 and S2 signals.  At energies above $\simeq 40$ keV, the efficiency drops sharply because of cuts to suppress the electronic-recoil contribution and to remove incidental light emission from defective photomultipliers.  In between, the efficiency is nearly constant at a value $\simeq 0.85$.

In addition to the neutrino-induced backgrounds shown in Fig.~\ref{fig:rates}, we take into account other types of detector backgrounds.  The most important is neutron-nucleus scattering events~\cite{Aalbers:2016jon, Pattavina:2020cqc, XENON:2020kmp}.  In XENON1T, this background after mitigation measures is comparable to that induced by atmospheric neutrinos.  It is caused by neutrons from the spontaneous fission of ${}^{232}$Th, ${}^{238}$U, and ${}^{235}$U, plus from secondary neutrons following $(\alpha, n)$ processes~\cite{XENON:2015gkh}.  For future xenon detectors, increasingly strong neutron-rejection measures will be needed for dark-matter searches, which will help DSNB searches.

Figure~\ref{fig:spectrum} shows realistic DSNB detection prospects for the example of XENON1T.  A $10^3 \times$ DSNB flux  (comparable to the present limit~\cite{Lunardini:2008xd}) is in range of detection.  Although Fig.~\ref{fig:spectrum} gives the impression that the peak of the DSNB spectrum is being probed, this recoil spectrum is heavily shaped by the detector efficiency.  The true range of the neutrino spectrum being probed can be estimated from Fig.~\ref{fig:Normalized_N}.

To isolate the DSNB from backgrounds, it is important to define a region of interest (ROI) in the energy spectrum.  We choose 4--15 keV to minimize the effects of backgrounds.  The choice of the low-energy edge of the ROI is particularly important, as the solar-neutrino backgrounds rise sharply with decreasing energy.  Within this ROI, the expected signal counts (for a DSNB enhancement factor of $10^3$, reflecting the present limit) are 0.47~yr$^{-1}$, while the expected background counts are 0.02~yr$^{-1}$ for the neutrino-induced backgrounds, and 0.03 yr$^{-1}$ for the total background.  (These values can be estimated from the plot by multiplying the height of the curve by the width of the ROI, $\Delta \ln E = 1.3$.)  The dominance of the signal (including the enhancement factor $10^3$) over the background is important, as explained in the next subsection, where we focus on how to probe even smaller DSNB fluxes with larger exposures.

We note that essentially the only useful improvements for detector sensitivity come from reducing the non-neutrino backgrounds (already assumed to be small) and increasing exposure.  Because of the neutrino backgrounds are irreducible, the ROI cannot be appreciably increased.  Of course, increasing the detector energy range will help measure the backgrounds, which is important~\cite{XENON:2020gfr}.  In principle, detectors with directional sensitivity could help reduce the solar-neutrino background~\cite{OHare:2015utx, Vahsen:2020pzb}, but the discriminating power would have to be excellent to make a difference because the solar background rises so steeply with decreasing energy.


\subsection{Calculated Flux Sensitivity}
\label{sec:resolution-statistics}

To calculate the sensitivity to the DSNB flux, we take into account two key points.  First, an ROI in the recoil spectrum can be defined that minimizes backgrounds. Second, the expected statistics are small.  It is then sufficient to compare the signal and background counts in the ROI, i.e., to use a one-bin analysis.  For each value of the detector exposure, we simulate many instances of a possible experiment, sampling the signal and background counts from Poisson distributions.  Assuming that the DSNB signal takes its fiducial value, no signal events are expected, and we would set an upper limit on the DSNB flux. We use a simple Feldman-Cousins treatment~\cite{Feldman:1997qc} to calculate the sensitivity for each simulation and the final limit as an average over the number of modeled experiments.  Ultimately, the sensitivity could be improved by using an unbinned maximum-likelihood approach, noting that the backgrounds are fully specified (our preliminary maximum-likelihood calculations yield very similar results to our approach above, which we maintain due to its physical clarity).

\begin{figure}[t]
\centering
\includegraphics[width=0.99\columnwidth]{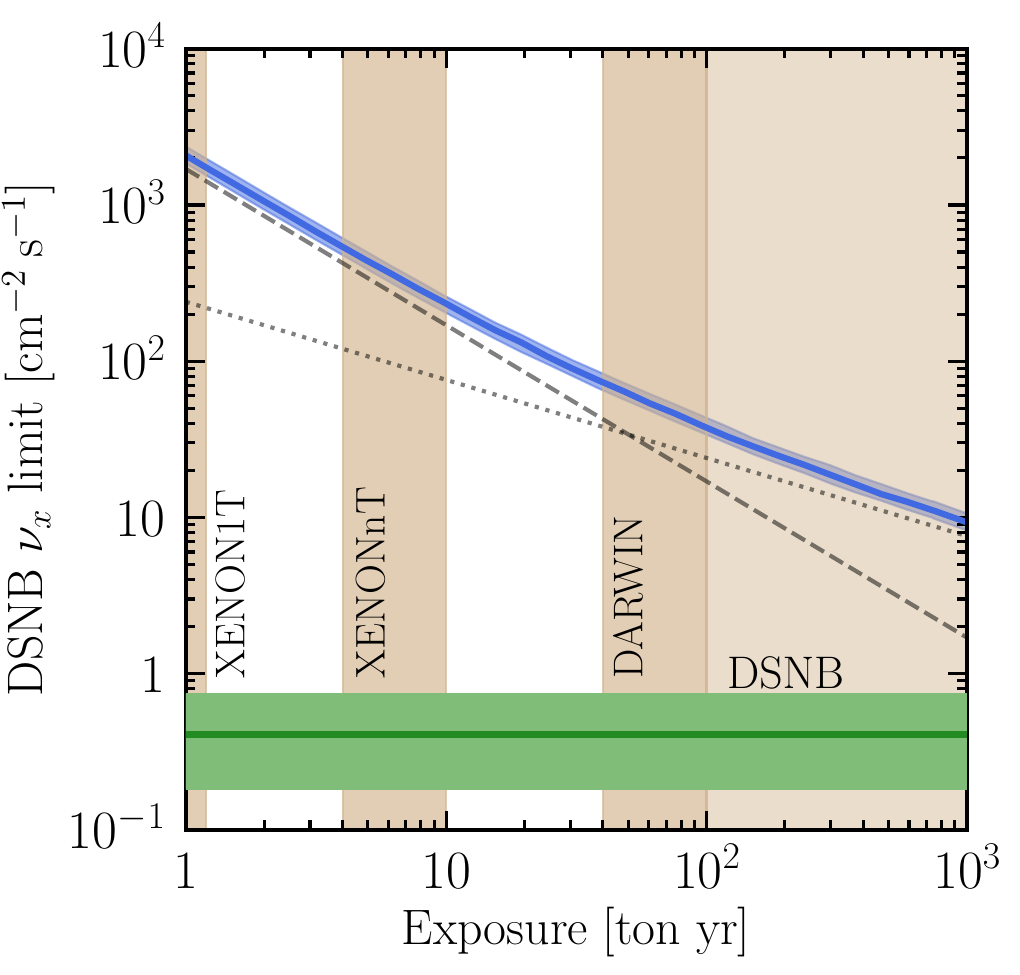}
\caption{
Scaling of the DSNB $\nu_x$ sensitivity with increasing exposure in xenon-based detectors (blue line) for an ROI of 4--15 keV.  The broken lines indicate scaling with exposure (dashed) and its square root (dotted).  The vertical bands show expected detector exposures.  The horizontal band shows our conservative DSNB modeling (fiducial model, with the band set by our minimal and maximal models).
}
\label{fig:limits}
\end{figure}

Figure~\ref{fig:limits} shows the projected 90\% C.L. upper limits on the DSNB $\nu_x$ flux (for each flavor) change with exposure for the series of xenon-based detectors.  For each detector (except for XENON1T, which is completed), a vertical band indicates an exposure range based on 1--2.5 years of runtime.  While we do not perform calculations for LUX-ZEPLIN~\cite{LZ:2019sgr}, it should have very similar capabilities to XENONnT~\cite{XENON:2020kmp}.  Our main focus is DARWIN~\cite{Aalbers:2016jon}, which will join the XENON and LUX-ZEPLIN collaborations. Given DARWIN's capabilities (including for detecting a supernova neutrino burst~\cite{Lang:2016zhv, Khaitan:2018wnf, Raj:2019wpy, Raj:2019sci}) and expense, it is conceivable that it would run for 25 years, which we indicate by a fainter extension of its band.  This figure shows that the series of xenon-based detectors can significantly improve sensitivity to DSNB $\nu_x$, despite the difficulties discussed above.  Even with present data from XENON1T, the sensitivity is comparable to the present limit using neutrino-electron scattering in SK~\cite{Lunardini:2008xd}.  Most important, the sensitivity in xenon-based detectors can improve by orders of magnitude.  While these detectors cannot reach standard predictions, testing if the DSNB $\nu_x$ flux is larger than expected would be valuable.

\begin{figure}[t]
\centering
\includegraphics[width=0.99\columnwidth]{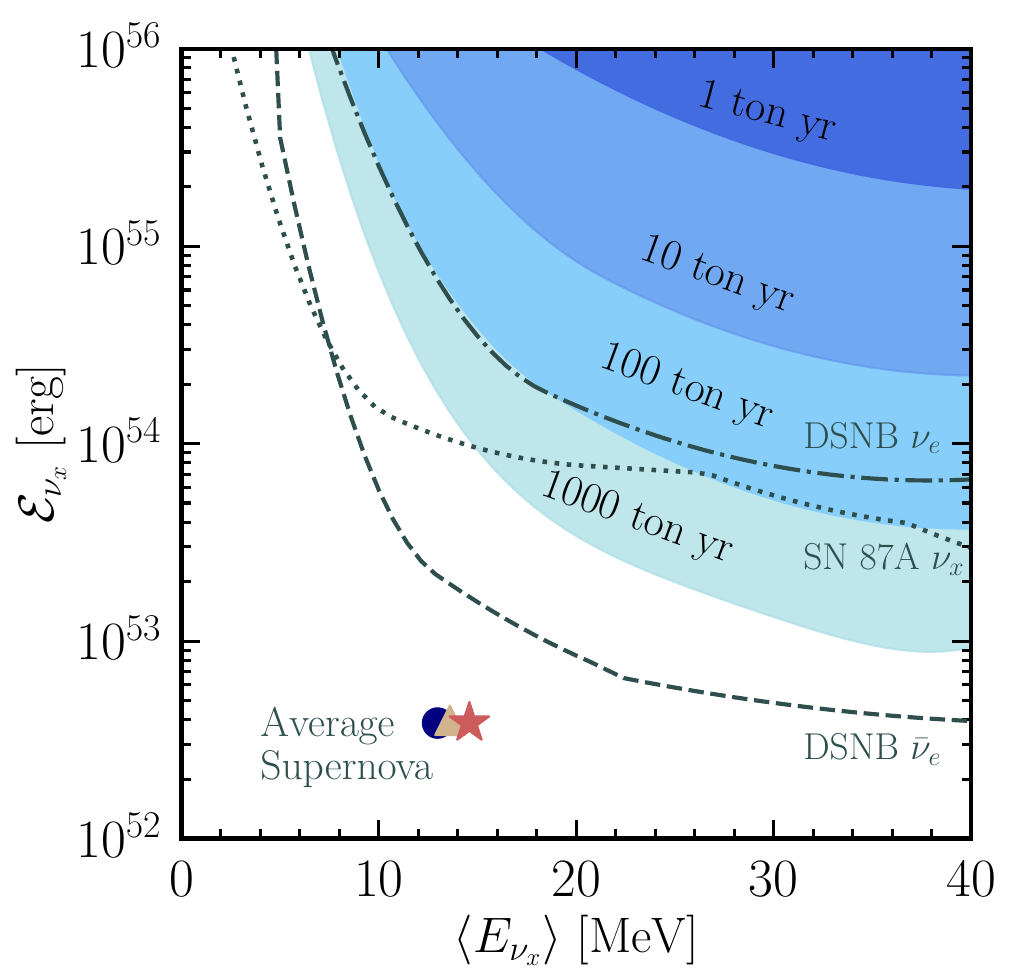}
\caption{
Sensitivity to DSNB $\nu_x$ (projected 90\% C.L. upper limits) in xenon-based detectors, characterized by the emitted energy (one of four flavors) and average energy (assumed common).  For comparison, we show the present SK limit on DSNB $\bar{\nu}_e$~\cite{Super-Kamiokande:2021acd}, the SNO limit on DSNB $\nu_e$~\cite{Beacom:2005it, Aharmim:2020agi}, and our SN 1987A limit on $\nu_x$ (see App.~\ref{app:SN1987A}), as well as three points that indicate our average emission per collapse in the fiducial, minimal, and maximal DSNB models (see App.~\ref{app:DSNB}).  
}
\label{fig:contour}
\end{figure}

For small exposures, the DSNB flux sensitivity scales inversely with exposure, because the number of background events in the ROI is $\ll 1$.  For large exposures, the sensitivity scales inversely with the square root of exposure, because the backgrounds are no longer negligible.  For large exposures, we found that the sensitivity can be improved by adjusting the ROI to keep the background contributions small.  This is an indication that the analysis should ultimately switch to using an unbinned maximum-likelihood approach.  Our results are insensitive to the details of the predicted DSNB spectrum.  The green band indicating the range of predictions (as in Figs.~\ref{fig:flux}, \ref{fig:rates}, and \ref{fig:spectrum}) is also shown here, where it is narrow because our results primarily probe the flux normalization, and not the spectrum shape.  For XENON1T (completed), the limit presented in Fig.~\ref{fig:limits} was calculated for energy threshold of~4~keV; employing a 4.9~keV threshold, as it is done for dark matter searches~\cite{Aprile:2018dbl}, yields a 20\% weaker limit, $2510$~cm$^{-2}$~s$^{-1}$.

As the exposure increases, the uncertainties on the predicted atmospheric neutrino backgrounds will become more important. At present, these are 20--30\% for the relevant neutrino energy range~\cite{Battistoni:2002ew, Honda:2011nf}.  To conservatively estimate the effects of including this, we did a one-bin maximum-likelihood analysis including Gaussian pull terms~\cite{OHare:2020lva}, assuming a 30\% atmospheric flux normalization uncertainty.  At the largest exposures we consider, this leads to a decrease in the flux sensitivity by approximately a factor of two.  We anticipate that the ultimate effects will be much less.  By the time the DARWIN exposure increases, these uncertainties will be reduced due to improved measurements in large neutrino detectors, as well as in DARWIN itself at recoil energies well above our ROI.  The very different shapes of the recoil spectra induced by the DSNB and by atmospheric neutrinos will allow good separation of these components in a full unbinned maximum-likelihood analysis.

Figure~\ref{fig:contour} shows the projected sensitivity of xenon-based detectors to the fundamental $\nu_x$ emission parameters: the energy emitted per single $\nu_x$ and the average energy.  To calculate these results, we redid our analysis using a simple DSNB model for which all supernovae emit the same thermal neutrino spectrum (see App.~\ref{app:DSNB}).  This figure, like Fig.~\ref{fig:limits}, shows that xenon-based detectors can significantly improve sensitivity to DSNB $\nu_x$, testing the possibility of nonstandard $\nu_x$ emission.  In addition, it demonstrates that large exposures of $\gtrsim 100$~ton~yr could improve upon the limits on $\nu_x$ from SN 1987A (see details in App.~\ref{app:SN1987A}) and the current sensitivity to DSNB $\nu_e$.  Our analysis focuses on the four $\nu_x$ flavors; once the sensitivity to those improves enough, the analysis should take into account that $\nu_e$ could also contribute to the signal.  While Fig.~\ref{fig:contour} extends to mean energies significantly higher than found by state-of-the-art supernova simulations~\cite{Janka:2016fox, OConnor:2018sti, Burrows:2020qrp}, it could be relevant for new-physics scenarios~\cite{Dolgov:2000jw, Ando:2003ie, Fogli:2004gy, Goldberg:2005yw, Baker:2006gm, Raffelt:2011nc, Farzan:2014gza, Arguelles:2016uwb, Jeong:2018yts, Suliga:2019bsq, Syvolap:2019dat, DeGouvea:2020ang, Mastrototaro:2019vug}.


\section{Discussion and Conclusions}
\label{sec:conclusions}

Understanding core-collapse supernovae is critical to progress in astrophysics and physics.  The key to progress is measuring their neutrino emission, which carries away 99\% of the released energy and which reflects the extreme, dynamic conditions in the core.  The better we understand the astrophysics of supernovae, the better we can probe neutrino physics by exploiting these extreme conditions.  And the better we understand neutrino physics, the better we can search for physics beyond the standard model, e.g., the emission of new particles.

From SN 1987A, we detected only the $\bar{\nu}_e$ flavor.  This makes it difficult to test the most basic prediction about the neutrino emission, that each flavor should carry away a comparable fraction of the total energy, and with comparable spectra.  And, even if this should be confirmed by the detection of the next Milky Way supernova, we still will not know how varied is the neutrino emission from the supernova population.  The  detection of DSNB $\bar{\nu}_e$ is coming within reach of SK~\cite{Super-Kamiokande:2021acd}, and there is good progress on DSNB $\nu_e$~\cite{Beacom:2005it, Aharmim:2020agi, Moller:2018kpn, Zhu:2018rwc}, but the limit for DSNB $\nu_x$, based on neutrino-electron scattering in SK, is very weak, $\sim 10^3 \; \mathrm{cm}^{-2}~\mathrm{s}^{-1}$ per flavor~\cite{Lunardini:2008xd}.
 
We show that DSNB $\nu_x$ can be well probed with xenon-based detectors for dark matter.  The advantages that make this possible are the large cross section for neutrino-nucleus coherent scattering, the very low backgrounds, and the plans to greatly increase the exposure.  With XENON1T (completed), one could probe fluxes comparable to the present limit.  With XENONnT/LUX-ZEPLIN (running) and especially DARWIN (proposed), great improvements could be made, reaching a sensitivity $\sim 10 \; \mathrm{cm}^{-2}~\mathrm{s}^{-1}$ per flavor.  While not reaching as far as needed, this sensitivity is enough to exclude many DSNB scenarios with new astrophysics or physics.  If the sensitivity can be pushed low enough, it could also improve sensitivity to DSNB $\nu_e$. 

Ultimately, to reach the goal of detecting the DSNB in all flavors~\cite{Lunardini:2008xd, Tabrizi:2020vmo}, even larger detectors will be needed.  We show in App.~\ref{app:lead} that the lead-based RES-NOVA-1, -2, -3 detectors ~\cite{Pattavina:2020cqc} have the potential to improve upon the sensitivity of xenon-based detectors, though work is needed to better assess the details.  For example, 2.5 years of RES-NOVA-3 could potentially improve on 2.5 years of DARWIN by about an order of magnitude.  Thus, though RES-NOVA may start later than the xenon-based detectors, it may be a path to detection of DSNB $\nu_x$.

Searches for DSNB $\nu_x$ could be complicated by the discovery of dark-matter scattering. But that fact would be adequate consolation!


\begin{acknowledgments}

We are grateful for helpful discussions with Baha Balantekin, Laura Baudis, Patrick Decowski, Ciaran O'Hare, Luca Pattavina, Troels Petersen, Louis Strigari, and especially Rafael Lang.  We thank the Garching Supernova group for access to their supernova-model data.  This work was supported by the Villum Foundation (Projects Nos.~13164 and 37358), the Danmarks Frie Forskningsfonds (Project No.~8049-00038B), the Deutsche Forschungsgemeinschaft through Sonderforschungbereich SFB 1258 ``Neutrinos and Dark Matter in Astro- and Particle Physics'' (NDM), and US National Science Foundation (Grant No.\ PHY-2012955 and No.\ PHY-2020275).

\end{acknowledgments}


\appendix


\section{Details on the DSNB model}
\label{app:DSNB}

Here we discuss our DSNB model in more detail; see also Ref.~\cite{Moller:2018kpn}.  The average flux per supernova in our three DSNB models (fiducial, plus minimal and maximal) corresponds to a weighted sum of fluxes from the NS- and BH-forming models.  This average flux for each $\nu_x$ flavor is characterized by a total energy, $\mathcal{E}_{\nu_x}$, and an average energy, $\langle E_{\nu_x}\rangle$. For our fiducial model, these are $0.40 \times 10^{53}~\mathrm{erg}$ and $13.6~\mathrm{MeV}$.  For our minimal model, they are $0.39 \times 10^{53}~\mathrm{erg}$ and $12.3~\mathrm{MeV}$; for our maximal model, they are $0.39 \times 10^{53}~\mathrm{erg}$ and $14.6~\mathrm{MeV}$.  These values are within the range expected from calculations, simulations, and observation~\cite{Jegerlehner:1996kx, Janka:2017vlw, OConnor:2018sti, Burrows:2020qrp}, though this is a relatively conservative prediction; the $\nu_x$ average energy is very close to that of the other flavors.

\begin{figure}[t]
\centering
\includegraphics[width=0.99\columnwidth]{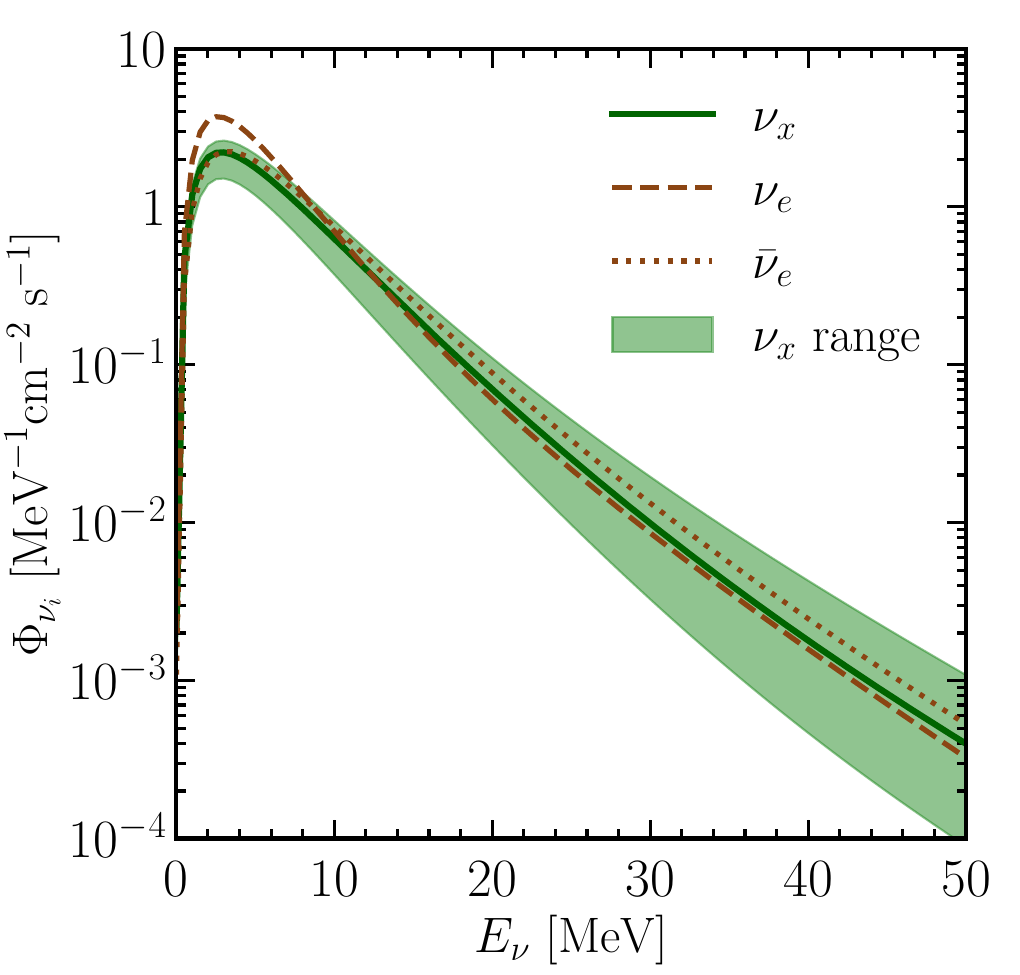}
\caption{
Predicted DSNB spectra.  The solid green line is for one of the four $\nu_x$ flavors in our fiducial model, with the band set by our minimal and maximal models (allowed by varying $f_\mathrm{BH}$, $R_{SN}(z=0)$, and the BH-forming model).  We also show the spectra of $\nu_e$ and $\bar{\nu}_e$ in our fiducial model.
}
\label{fig:DSNB_nue_vs_nux}
\end{figure}

Figure~\ref{fig:DSNB_nue_vs_nux} shows the DSNB spectra in our fiducial model, with an uncertainty range indicated for $\nu_x$ alone.  The variation at lower energies is dominated by the local supernova rate, $R_\mathrm{SN}(z=0)$, while that at higher energies by the BH-forming fraction, $f_\mathrm{BH}$.  We expect a hierarchy of average energies in the different flavors due to their different interactions with matter, with the $\nu_e$, $\bar{\nu}_e$, and $\nu_x$ spectra being progressively hotter.  This is seen for $\nu_e$, but for $\bar{\nu}_e$ there is a subtlety.
A local temperature maximum emerges behind the free-streaming radius for $\nu_x$ but ahead of the one $\bar{\nu}_e$~\cite{Mirizzi:2015eza}, so the $\nu_x$ lose additional energy by scattering, lowering their average energy.


\section{SN 1987A Limits on $\nu_x$ Emission}
\label{app:SN1987A}

Here we estimate upper limits on the supernova $\nu_x$ emission from SN 1987A observations.  To summarize the context, three detectors registered events consistent with charged-current $\bar{\nu}_e$ interactions with free (hydrogen) protons.  The water-based Kamiokande-II (KamII) detector observed eleven events~\cite{PhysRevLett.58.1490}.  The water-based Irvine-Michigan-Brookhaven (IMB)~\cite{Bionta:1987qt} detector observed eight events.  The scintillator-based Baksan Neutrino Observatory (Baksan)~\cite{ALEXEYEV1988209} observed five events (most, but not all, were likely due to detector backgrounds).  However, these detectors were capable, in principle, of observing $\nu_x$ events via neutral-current interactions.  The yields are suppressed by smaller cross sections, higher interaction thresholds, or because they produce observable energies below the detection thresholds.  We estimate the $\nu_x$ limits using three detection channels: neutrino elastic scattering with electrons ($\nu_x + e^{-}$)~\cite{Giunti:2007ry}, neutrino excitation of oxygen with gamma-ray emission ($\nu_x + \mathrm{O} $)~\cite{Langanke:1995he, Beacom:1998ya}, and the same for carbon ($\nu_x + \mathrm{C}$)~\cite{199815, Laha:2013hva, Lu:2016ipr}.

For the first two detection channels, the strongest limits come from considering the Kam-II detector.  The IMB detector had too high of an effective energy threshold for both channels and the Baksan detector was too small for the first.  Kam-II was a $\sim$ 2-kton water Cherenkov detector with a detection threshold energy of 7.5 MeV (electron total energy) and an energy-dependent efficiency~\cite{PhysRevLett.58.1490}.  For $\nu_x + \mathrm{O}$ scattering, we take the approximate energy variation of the cross section from Ref.~\cite{Beacom:1998ya}, assume the branching ratios from Ref.~\cite{Langanke:1995he} (even though these should vary with energy), include only gamma rays above 9 MeV, and assume that these Compton-scatter electrons with (on average) 80\% of the gamma-ray energy.  We make conservative choices for the expected numbers of events required for clear $\nu_x$ detection.  For $\nu_x + e^{-}$ scattering, we require a Poisson mean of at least 4.0 events, so that there is a 90\% probability of observing more than one event (as one Kam-II event was forward).  For $\mathrm{O} + \nu_x$ scattering, we require a Poisson mean of at least 6.8 events, so that there is a 90\% probability of observing more than three events (as a few Kam-II events were in this energy range).  For the third detection channel, we calculate the limits assuming the Baksan detector, which had a mass of $\sim 200$ ton of scintillator.  It had a detection threshold energy of 8~MeV and an efficiency of 80\% above 12 MeV \cite{ALEXEYEV1988209}.  The de-excitation of carbon is via a gamma ray of energy 15.11 MeV~\cite{199815}.  We conservatively require that the total $\nu_x$ flux yields more than 6.8 events because a few Baksan events were in this energy range. 

\begin{figure}[t]
\centering
\includegraphics[width=0.99\columnwidth]{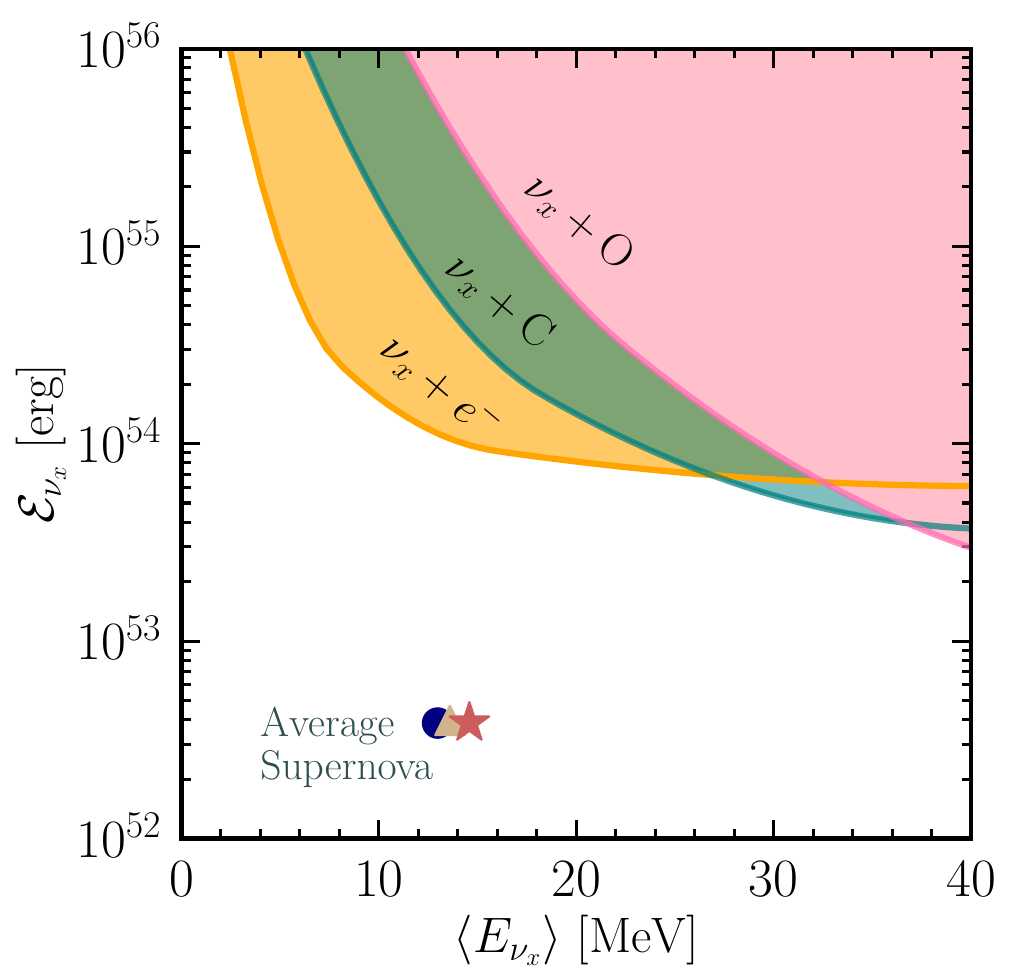}
\caption{
Limits on supernova $\nu_x$ emission from SN 1987A.  The results are estimated from data from Kam-II ($\nu_x + e^{-}$ and $\nu_x + \mathrm{O}$) and Baksan ($\nu_x + \mathrm{C}$), characterized by total energy and average energy for each flavor of $\nu_x$.
}
\label{fig:SN1987A}
\end{figure}

Figure~\ref{fig:SN1987A} shows the limits on supernova $\nu_x$ emission calculated from SN 1987A, which may or may not be a typical supernova.  The most robust limit is that based on neutrino-electron scattering.  Its variation with average energy can be understood simply.  At high average energies, the limit is flat because the cross section rises with average energy while the number of neutrinos falls with average energy (assuming a fixed total energy).  At low average energies, the effects of the detector threshold are important.  The neutrino-oxygen and neutrino-carbon limits are more approximate and might be improved by further work.  Their variation with average energy reflects not only the effects of detector thresholds, but also the steeper energy dependence of the cross sections (approximately quartic in the energy above threshold for oxygen~\cite{Langanke:1995he, Beacom:1998ya} and approximately quadratic in the energy above threshold for carbon~\cite{199815, Laha:2013hva, Lu:2016ipr}).

\section{Results for Lead-Based Detectors}
\label{app:lead}

Here we summarize results for lead-based detectors, for which the calculations are very similar to those for xenon-based detectors.  Our approach is to assess the best possible sensitivity, testing if this is promising.  We consider RES-NOVA, which would use archaeological lead as an active detector volume~\cite{Pattavina:2020cqc}. The current proposal has three phases, RES-NOVA-1, -2, and -3, with effective masses of 2.4, 31, and 456 ton, respectively, where we assume pure lead for now.  These masses are substantially larger than assumed for the xenon-based detectors.

\begin{figure}[t]
\centering
\includegraphics[width=0.99\columnwidth]{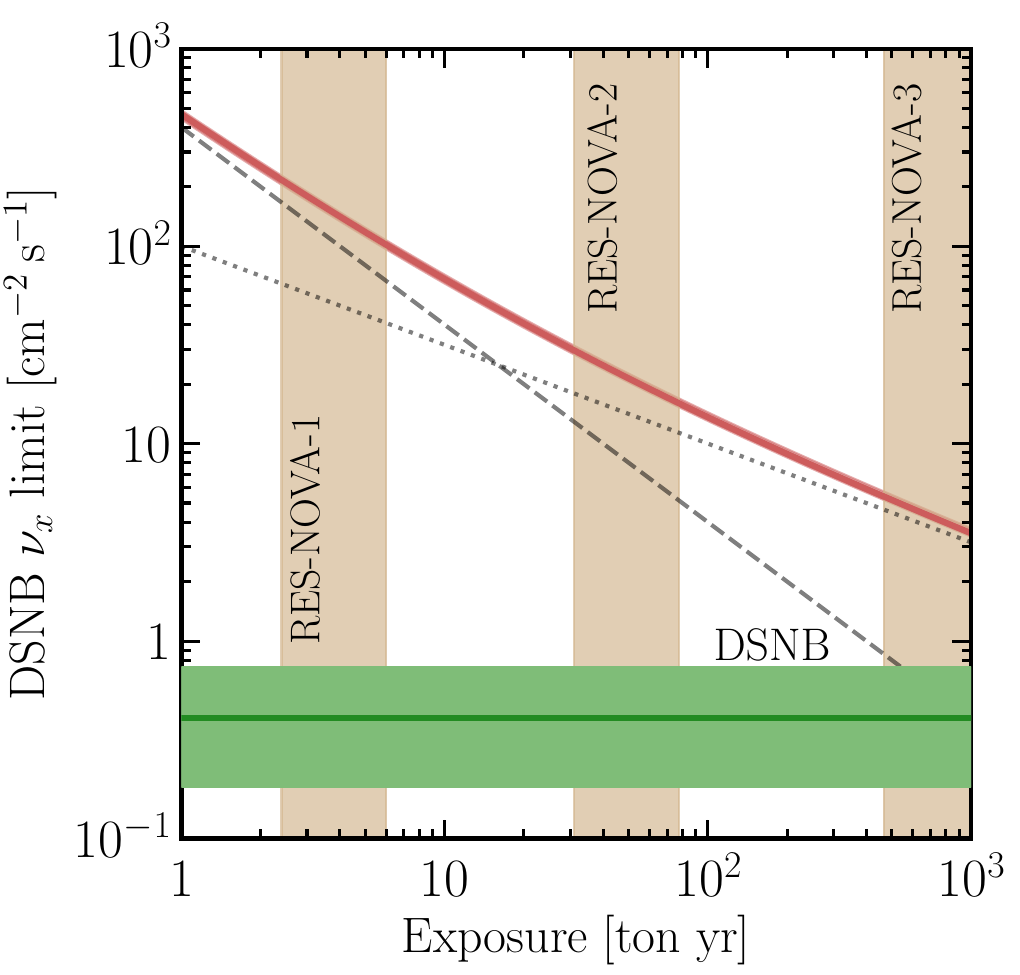}
\caption{
Scaling of the best possible DSNB $\nu_x$ sensitivity with increasing exposure in pure lead-based detectors (solid red line) for an ROI of 2--15 keV.  The broken lines indicate scaling with exposure and its square root.  The vertical bands show expected exposures.  The horizontal band shows our conservative DSNB modeling (fiducial model, with the band set by our minimal and maximal models).  RES-NOVA-3 might reach much greater exposures than shown.  Compare to Fig.~\ref{fig:limits} for xenon-based detectors.
}
\label{fig:limits-Pb}
\end{figure}

The detector backgrounds for RES-NOVA have been evaluated for the case of supernova neutrino burst detection~\cite{RES-NOVA:2021gqp, Pattavina:2020cqc}.  There are large background rates due to the decay chains of $^{238}\mathrm{U}$, $^{210}\mathrm{Pb}$, and $^{232}\mathrm{Th}$ in the detector. These rates surpass the event rates induced by solar, atmospheric, and DSNB neutrinos by a few orders of magnitude, but it is anticipated that they can be greatly reduced by particle identification techniques~\cite{Beeman:2012wz, RES-NOVA:2021gqp}.  Further studies are needed to calculate the background rates for DSNB searches.

To estimate the best possible signal sensitivity, we consider only solar- and atmospheric-neutrino backgrounds (in particular, neglecting neutron backgrounds).  Further, we assume 100\% efficiency and neglect energy-resolution smearing. In Ref.~\cite{RES-NOVA:2021gqp}, the energy resolution was assumed to be 0.2~keV, independent of the recoil energy, which would have minimal effects on our results.  We find that an ROI of 2--15 keV works well.

If the detector is realized with lead crystals, for example PbWO$_4$~\cite{Pattavina:2020cqc}, there are additional issues.  First, the total mass of lead targets would be smaller by about 30\%, with the difference being made up by the other elements.  Second, and more important, these other elements become important targets for neutrinos.  For oxygen in particular, its lower mass allows a larger recoil energy for the same neutrino energy.  This has the effect of moving the solar-neutrino background to higher recoil energies, overwhelming the DSNB signal until almost 55 keV.  In this case, the RES-NOVA sensitivity to DSNB $\nu_x$ worsens by a factor of about $\sim 30$, which would not be competitive.  Thus it would be important to either use pure lead or to use crystals based on only heavy elements.

\begin{figure}[t]
\centering
\includegraphics[width=0.99\columnwidth]{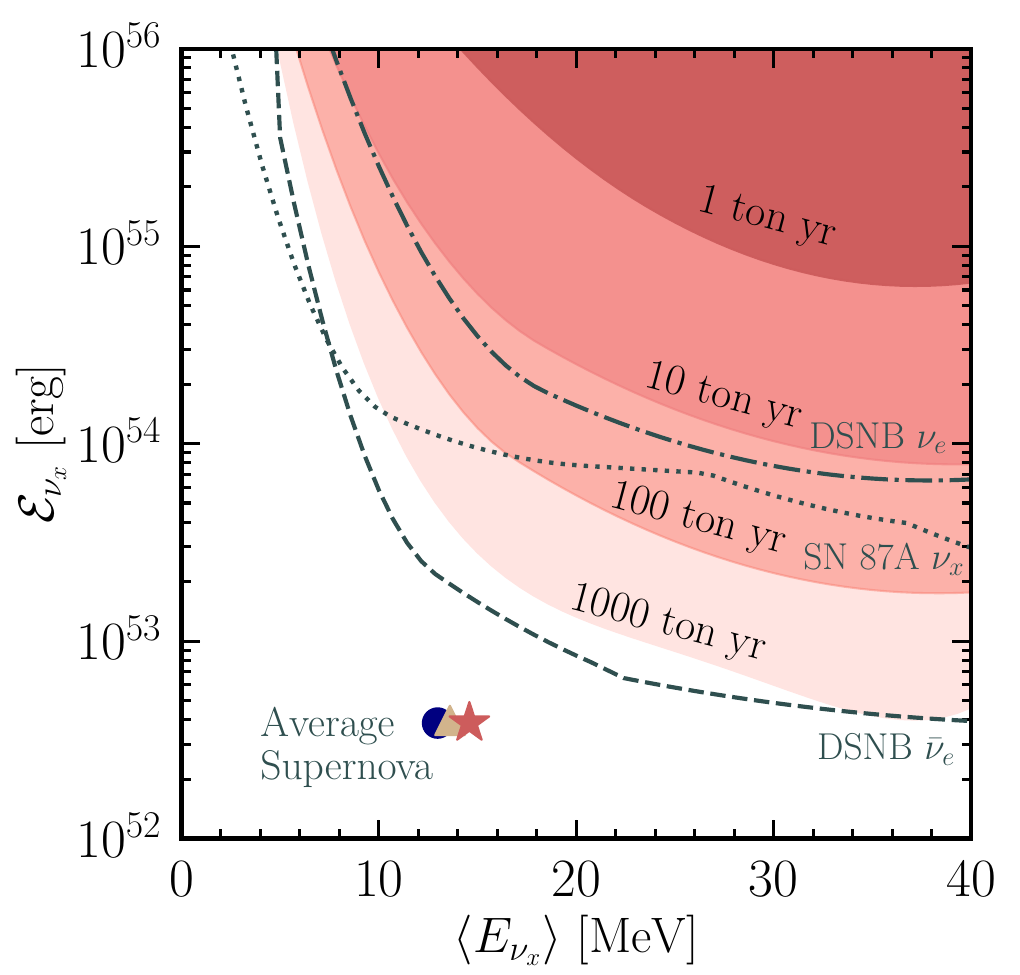}
\caption{
Sensitivity to DSNB $\nu_x$ (projected 90\% C.L. upper limits) in pure lead-based detectors, characterized by the emitted energy (one of four flavors) and average energy (assumed common).  For comparison, we show the present SK limit on DSNB $\bar{\nu}_e$~\cite{Super-Kamiokande:2021acd}, the SNO limit on DSNB $\nu_e$~\cite{Beacom:2005it, Aharmim:2020agi}, and our SN 1987A limit on $\nu_x$ (see App.~\ref{app:SN1987A}), as well as three points that indicate our average emission per collapse in the fiducial, minimal, and maximal DSNB models (see App.~\ref{app:DSNB}).  Compare to Fig.~\ref{fig:contour} for xenon-based detectors.
}
\label{fig:contour-Pb}
\end{figure}

Figure~\ref{fig:limits-Pb} shows our calculated flux sensitivity for DSNB $\nu_x$ in pure lead-based detectors.  As for xenon-based detectors, the sensitivity would immediately be competitive and improve rapidly.  If the detector backgrounds are as small as hoped, the sensitivity could ultimately be better than for xenon-based detectors.

Figure~\ref{fig:contour-Pb} shows the projected sensitivity of pure lead-based detectors to the $\nu_x$ emission parameters.


\newpage
\phantom{i}
\newpage
\bibliography{DSNB}

\end{document}